# Estimating quantities conserved by virtue of scale invariance in timeseries

Erik D. Fagerholm[1,*], W.M.C. Foulkes[2], Yasir Gallero-Salas[3,4], Fritjof Helmchen[3,4], Karl J. Friston[5], Rosalyn J. Moran[1], Robert Leech[1]

[1] Department of Neuroimaging, King's College London
[2] Department of Physics, Imperial College London
[3] Brain Research Institute, University of Zürich
[4] Neuroscience Center Zürich
[5] Wellcome Trust Centre for Neuroimaging, University College London

* Corresponding author: erik.fagerholm@kcl.ac.uk

In contrast to the symmetries of translation in space, rotation in space, and translation in time, the known laws of physics are not universally invariant under transformation of scale. However, the action can be invariant under change of scale in the special case of a scale free dynamical system that can be described in terms of a Lagrangian, that itself scales inversely with time. Crucially, this means symmetries under change of scale *can* exist in dynamical systems under certain constraints. Our contribution lies in the derivation of a generalised scale invariant Lagrangian – in the form of a power series expansion – that satisfies these constraints. This generalised Lagrangian furnishes a normal form for dynamic causal models (i.e., state space models based upon differential equations) that can be used to distinguish scale invariance (scale symmetry) from scale freeness in empirical data. We establish face validity with an analysis of simulated data and then show how scale invariance can be identified – and how the associated conserved quantities can be estimated – in neuronal timeseries.

A symmetry is a transformation to a physical law that leaves its mathematical form invariant[1]. For instance, most of the known laws of physics are invariant under translation in space, rotation in space, and translation in time. In other words, we can state the following: having taken all influencing factors into account, it is impossible for an external observer to determine whether a dynamical system has been shifted to a new location, rotated by a fixed angle, or whether its onset has been shifted in time.

However, the laws are generally not invariant under transformation of scale. Richard Feynman famously described an intuitive example of why this is the case for a scale transformation within a gravitational field. He asked the audience to consider a thought experiment in which an intricate cathedral made of matchsticks was increased in size to the point where it would instead be made of great logs, thus collapsing under its own weight. The scale dependence of this system is further emphasized by the observation that:

*"…when you're comparing two things you must change everything that's in the system. The little cathedral made with matchsticks is attracted to the Earth. So, to make the comparison I should make the big cathedral attracted to an even bigger Earth. Too bad – a bigger Earth would attract it even more and the sticks would break even more surely."* [2]

Scale symmetries are therefore not as universally applicable as translation in space, rotation in space, and translation in time, but there are nevertheless known constraints under which scale symmetries *can* arise in dynamical systems. It is the purpose of the presented work to use these constraints in creating a method for estimating scale symmetries – and their associated conserved quantities – in empirical timeseries.

We proceed by introducing three key concepts used throughout this paper: a) Lagrangian mechanics; b) scale freeness; and c) scale invariance.

**Lagrangian mechanics:** Consider a system described by a Lagrangian with explicit time-dependence $\mathcal{L}(q, \dot{q}, t)$ to facilitate the analysis of driven systems. According to Hamilton's principle[3], the trajectory $q(t)$ followed by the system from any chosen initial point $q_i$ at time $t_i$ to any chosen final point $q_f$ at time $t_f$ renders the action, given by:

$$S[q(t)] = \int_{t_i}^{t_f} \mathcal{L}\left(q(t), \frac{dq(t)}{dt}, t\right) dt,$$ [1]

stationary.

In other words, for any infinitesimal path variation $\delta q(t)$ satisfying $\delta q(t_i) = \delta q(t_f) = 0$, we see that:

$$S[q(t) + \delta q(t)] = S[q(t)] + \mathcal{O}[(\delta q)^2].$$ [2]

One can then use standard arguments[4] to show that any trajectory $q(t)$ for which the action is stationary is a solution of the Euler-Lagrange equation:

$$\frac{\partial \mathcal{L}}{\partial q} - \frac{d}{dt}\left(\frac{\partial \mathcal{L}}{\partial \dot{q}}\right) = 0.$$ [3]

**Scale freeness:** Scale freeness describes a situation in which different levels of magnification of a dynamical system are indistinguishable to within a multiplicative constant[5]. Given the set of points $(t, q)$ lying on some chosen trajectory $q(t)$, we define the corresponding scaled trajectory as the set of points $(t_s, q_s) = (\lambda^\alpha t, \lambda q)$, where $\lambda \, (> 0)$ is a spatial scale factor. The time coordinate has been rescaled by $\lambda^\alpha$, where $\alpha$ is a system-dependent constant. The scaled trajectory passes through the point:

$$q_s = \lambda q,$$
$$t_s = \lambda^\alpha t,$$ [4]

implying that $q_s(\lambda^\alpha t) = \lambda q(t)$, or, equivalently that:

$$q_s(t_s) = \lambda q(\lambda^{-\alpha} t_s),$$ [5]

from which it follows that:

$$\frac{dq_s(t_s)}{dt_s} = \lambda^{1-\alpha} \dot{q}(\lambda^{-\alpha} t_s).$$ [6]

We refer to the system's dynamics as being 'scale free' if, for any path $q(t)$:

$$S[q(t)] = \kappa \, S[q_s(t_s)], \qquad\qquad [7]$$

where $\kappa$ is a constant that may depend on the scale factor $\lambda$ but is independent of path.

More explicitly, using [1] and [7], we see that the system is scale free if:

$$\int_{t_i}^{t_f} \mathcal{L}\left(q(t), \frac{dq(t)}{dt}, t\right) dt = \kappa \int_{\lambda^\alpha t_i}^{\lambda^\alpha t_f} \mathcal{L}\left(q_s(t_s), \frac{dq_s(t_s)}{dt_s}, t_s\right) dt_s. \qquad [8]$$

Assuming that $q(t)$ is a physical trajectory derived by applying Hamilton's principle to a scale free action, it follows from [2] and [7] that:

$$S[q_s(t_s) + \delta q_s(t_s)] = S[q_s(t_s)] + O[(\delta q_s)^2], \qquad\qquad [9]$$

for all infinitesimal path variations $\delta q_s(t_s)$.

This shows that the scaled path described by [5] also renders the action stationary, i.e. if $q(t)$ is a possible physical trajectory then the same can be said for the scaled trajectory $q_s(t_s) = \lambda q(\lambda^{-\alpha} t_s)$.

Scale freeness has been observed in a variety of physical and biological settings[6]. These include neural systems across different species[7-9], in which scale freeness is identified by signatures of critical neuronal dynamics[10,11], and is considered to offer functional[12], developmental[13], as well as evolutionary[14,15] advantages.

**Scale invariance:** We say that a system is 'scale invariant', or equivalently 'scale symmetric', if it is impossible to determine the magnification at which its evolution is observed. In other words, scale invariance means that a system is *perfectly* unchanged under transformation of scale, i.e. by setting $\kappa = 1$ in equation [8]:

$$\int_{t_i}^{t_f} \mathcal{L}\left(q(t), \frac{dq(t)}{dt}, t\right) dt = \int_{\lambda^\alpha t_i}^{\lambda^\alpha t_f} \mathcal{L}\left(q_s(t_s), \frac{dq_s(t_s)}{dt_s}, t_s\right) dt_s, \qquad [10]$$

This paper comprises three sections.

In the first, we show that an equation of motion leads to a scale invariant action under the constraint that its Lagrangian scales inversely with time. Using this constraint, the main contribution of this paper is presented via the derivation of a generalised scale invariant Lagrangian, which can be used to model timeseries from any scale free system that follows the principle of stationary action. We then use Noether's theorem to write the expression for the family of conservation laws associated with this generalised Lagrangian.

In the second, we demonstrate proof of principle by showing that the generalised Lagrangian can be used to identify scale invariance in arbitrary timeseries. Specifically, we show that scale invariance can be distinguished from the less restrictive condition of scale freeness via simulations of a classical particle.

In the third, using murine calcium imaging and macaque monkey fMRI datasets, we show that neural systems support a neurobiologically-based quantity that is conserved by virtue of scale symmetries.

**The condition for scale invariance:** We see via [5], [6], and [10] that:

$$\int_{t_i}^{t_f} \mathcal{L}(q(t), \dot{q}(t), t) dt = \int_{\lambda^\alpha t_i}^{\lambda^\alpha t_f} \mathcal{L}(\lambda q(\lambda^{-\alpha} t_s), \lambda^{1-\alpha} \dot{q}(\lambda^{-\alpha} t_s), t_s) dt_s$$

$$= \lambda^\alpha \int_{t_i}^{t_f} \mathcal{L}(\lambda q(t), \lambda^{1-\alpha} \dot{q}(t), \lambda^\alpha t) dt, \qquad [11]$$

where, using [4], the integration variable on the right-hand side was changed from $t_s$ to $t = \lambda^{-\alpha} t_s$.

Since the path of integration is arbitrary, it follows that the action is scale invariant if and only if the Lagrangian satisfies:

$$\mathcal{L}_s(q, \dot{q}, t) \equiv \mathcal{L}(\lambda q, \lambda^{1-\alpha} \dot{q}, \lambda^\alpha t) = \lambda^{-\alpha} \mathcal{L}(q, \dot{q}, t), \qquad [12]$$

where the identity defines the scaled Lagrangian $\mathcal{L}_s$ and the equality describes the condition for scale invariance.

We therefore see that scale invariance can exist in scale free systems if these can be described by a Lagrangian that scales inversely with time. In other words, given a spatiotemporal transformation in which $q \to \lambda q$ and $t \to \lambda^\alpha t$, the system is scale invariant if the Lagrangian transforms as $\mathcal{L} \to \lambda^{-\alpha}\mathcal{L}$, implying that $\mathcal{L}$ scales as $1/t$.

**A family of scale invariant Lagrangians:** Here, we present the contribution of this paper via the derivation of a generalised scale invariant Lagrangian that can be used to identify scale invariance in timeseries from any scale free dynamical system following the principle of stationary action.

We can write an expression for a Lagrangian $\mathcal{L}(q, \dot{q}, t)$ as a sum over power terms:

$$\mathcal{L}(q, \dot{q}, t) = \sum_{x,y,z} C_{xyz} q^x \dot{q}^y t^z, \qquad [13]$$

where $x$, $y$ and $z$ are constants and $C_{xyz}$ is an arbitrary expansion coefficient.

Using [12] we see that [13] is scale invariant if:

$$\mathcal{L}_s(q, \dot{q}, t) = \mathcal{L}(\lambda q, \lambda^{1-\alpha}\dot{q}, \lambda^\alpha t) = \sum_{x,y,z} \lambda^{x+(1-\alpha)y+\alpha z} C_{xyz} q^x \dot{q}^y t^z = \lambda^{-\alpha}\mathcal{L}(q, \dot{q}, t), \qquad [14]$$

and since $\lambda$ is arbitrary, this implies that:

$$x + (1-\alpha)y + \alpha z = -\alpha, \qquad [15]$$

$\forall\, x, y, z : C_{xyz} \neq 0$.

Note that one can in principle restrict the allowed values of the exponents in [14] to positive integers in order to obtain an analytic function. However, for the purpose of the work presented here we do not place such a restriction, so as to allow for greater flexibility in subsequent timeseries analyses. We can then uniquely determine the value of $x$ via [15], given that $\alpha$ is known and that a non-zero term with specific values of $y$ and $z$ exists. This in turn means that we can replace the triple summation in [13] with a double summation:

$$\mathcal{L}(q, \dot{q}, t) = q^{-\alpha} \sum_{y,z} C_{yz} q^{y(\alpha-1)-z\alpha} \dot{q}^y t^z, \qquad [16]$$

which describes a family of scale invariant Lagrangians.

**Noether's theorem and scale symmetry:** In 1918 Noether demonstrated that for every continuous symmetry of the action of a dynamical system there exists a corresponding conservation law[16]. This theorem tells us that it is by virtue of the symmetries of translation in space, rotation in space, and translation in time that the corresponding quantities of linear momentum, angular momentum, and energy are conserved, respectively. We proceed by reminding the reader of how Noether's theorem can be used to derive an expression for the quantity that is conserved by virtue of a symmetry under transformation of scale in space and time.

Beginning from the statement of scale invariance [11], we set $\lambda = 1 + \epsilon$, where $\epsilon$ is an arbitrarily small constant. This allows for any scale transformation to be constructed by sequentially applying such infinitesimal transformations.

Working to first order in $\epsilon$ we can write [11] as follows:

$$\int_{t_i}^{t_f} \mathcal{L}(q, \dot{q}, t) dt = \int_{t_i}^{t_f} \mathcal{L}\big((1+\epsilon)q, (1+(1-\alpha)\epsilon)\dot{q}, (1+\alpha\epsilon)t\big)(1+\alpha\epsilon) dt. \qquad [17]$$

Expanding the right-hand side and cancelling the $\epsilon$-independent terms we see that:

$$\epsilon \int_{t_i}^{t_f} \left\{ q \frac{\partial \mathcal{L}}{\partial q} + (1-\alpha)\dot{q} \frac{\partial \mathcal{L}}{\partial \dot{q}} + \alpha t \frac{\partial \mathcal{L}}{\partial t} + \alpha \mathcal{L} \right\} dt = 0, \qquad [18]$$

and since $\frac{d\mathcal{L}}{dt} = \frac{\partial \mathcal{L}}{\partial t} + \frac{\partial \mathcal{L}}{\partial q} \dot{q} + \frac{\partial \mathcal{L}}{\partial \dot{q}} \ddot{q}$, this is equivalent to:

$$\epsilon \int_{t_i}^{t_f} \left\{ q \frac{\partial \mathcal{L}}{\partial q} + (1-\alpha)\dot{q} \frac{\partial \mathcal{L}}{\partial \dot{q}} + \alpha \mathcal{L} + \alpha t \left( \frac{d\mathcal{L}}{dt} - \frac{\partial \mathcal{L}}{\partial q} \dot{q} - \frac{\partial \mathcal{L}}{\partial \dot{q}} \ddot{q} \right) \right\} dt = 0. \qquad [19]$$

If we now stipulate that $q(t)$ is a physical path of the system, we can use the Euler-Lagrange equation [3] to eliminate $\frac{\partial \mathcal{L}}{\partial q}$ from [19] to obtain:

$$\epsilon \int_{t_i}^{t_f} \left\{ (q - \alpha t \dot{q}) \frac{d}{dt} \left( \frac{\partial \mathcal{L}}{\partial \dot{q}} \right) + (1 - \alpha) \dot{q} \frac{\partial \mathcal{L}}{\partial \dot{q}} + \alpha \mathcal{L} + \alpha t \frac{d\mathcal{L}}{dt} - \alpha t \ddot{q} \frac{\partial \mathcal{L}}{\partial \dot{q}} \right\} dt = 0, \quad [20]$$

which can be rewritten as:

$$\epsilon \int_{t_i}^{t_f} \frac{d}{dt} \left\{ (q - \alpha t \dot{q}) \frac{\partial \mathcal{L}}{\partial \dot{q}} + \alpha t \mathcal{L} \right\} dt = 0, \quad [21]$$

from which we see that the quantity:

$$N = (q - \alpha t \dot{q}) \frac{\partial \mathcal{L}}{\partial \dot{q}} + \alpha t \mathcal{L} = \left( \mathcal{L} - \dot{q} \frac{\partial \mathcal{L}}{\partial \dot{q}} \right) \alpha t + \frac{\partial \mathcal{L}}{\partial \dot{q}} q, \quad [22]$$

must have the same value at the (arbitrary) initial and final times $t_i$ and $t_f$.

We therefore arrive at a special case of Noether's theorem in [22] applicable to scale invariant systems.

**Conservation laws associated with a family of scale invariant Lagrangians:** Using [22] we can now write an expression for the conserved quantities associated with the family of scale invariant Lagrangians in [16]:

$$N = \alpha q^{-\alpha} t \sum (1 - y) C_{yz} q^{y(\alpha-1) - z\alpha} \dot{q}^y t^z + q^{1-\alpha} \dot{q}^{-1} \sum y C_{yz} q^{y(\alpha-1) - z\alpha} \dot{q}^y t^z. \quad [23]$$

which arise within a specific functional subset (Fig. 1):

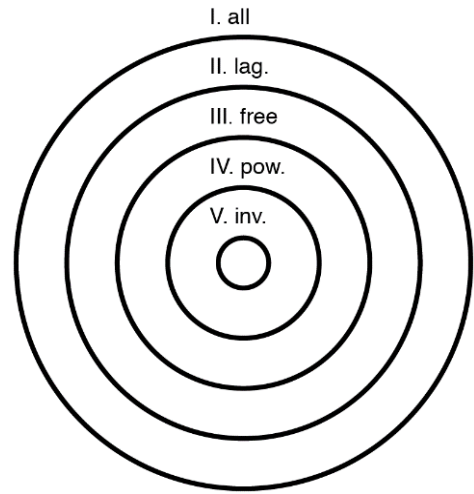

*Figure 1: scale invariant functions. In order of decreasing size, the areas of the circles represent the space of I. All possible functions; II. Functions that can be cast within a Lagrangian framework; III. Scale free Lagrangians, IV. Scale free Lagrangians that can be expressed as a power series; and V. Scale invariant power series Lagrangians. It is for this smallest subset of functions for which the quantities in [23] are conserved by virtue of scale invariance.*

**Free classical particle:** Here we analyse what is perhaps the simplest example of a dynamical system, in the form of a classical particle moving in the absence of a changing potential energy. The motion of this free particle is described by the following Lagrangian:

$$\mathcal{L} = \frac{1}{2}m\dot{q}^2, \qquad [24]$$

which, using [6], transforms under scale as follows:

$$\mathcal{L} \rightarrow \mathcal{L}_s = \lambda^{2(1-\alpha)}\frac{1}{2}m\dot{q}^2, \qquad [25]$$

which satisfies the condition for scale invariance in [12] when:

$$2(1-\alpha) = -\alpha \implies \alpha = 2, \qquad [26]$$

which, together with [22] and [24] show us that the corresponding conserved quantity is:

$$N = \left(\frac{1}{2}m\dot{q}^2 - m\dot{q}^2\right)2t + mq\dot{q} = m\dot{q}(q - \dot{q}t), \qquad [27]$$

and since for a free particle, $\dot{q}$ is constant and $q = q_i + \dot{q}(t - t_i)$, it becomes clear that $N$ is indeed conserved along the trajectory.

**Classical particle in a potential:** We now consider the effect of adding a potential energy term to [24]:

$$\mathcal{L} = \frac{1}{2}m\dot{q}^2 + kq^p, \qquad [28]$$

where $k$ and $p$ are constants.

Using [4] and [6] we see that [28] transforms under scale as:

$$\mathcal{L} \rightarrow \mathcal{L}_s = \lambda^{2(1-\alpha)}\frac{1}{2}m\dot{q}^2 + \lambda^p kq^p. \qquad [29]$$

From [12] we know that a scale symmetry exists if and only if $\mathcal{L}_s = \lambda^{-\alpha}\mathcal{L}$, implying that $2(1 - \alpha) = p = -\alpha$ and hence that $\alpha = 2$ and $p = -2$. To be scale invariant, the potential must therefore be an inverse square and the Lagrangian in [28] must take the form:

$$\mathcal{L} = \frac{1}{2}m\dot{q}^2 + kq^{-2}. \qquad [30]$$

Or in other words, in order for a particle described by [28] to be invariant under change of scale, it must be acted upon by a force that varies inversely as the cube of position – a special case that has been analysed previously[17,18].

Using [22] and [30] we see that:

$$N = mq\dot{q} - (m\dot{q}^2 - 2kq^{-2})t, \qquad [31]$$

which can be simplified by noting from [30] that the system's total energy, or Hamiltonian, is $\mathcal{H} = \frac{1}{2}m\dot{q}^2 - kq^{-2}$, meaning that [31] can be re-written as:

$$N = mq\dot{q} - 2\mathcal{H}t, \qquad [32]$$

hence giving us an expression for the quantity that is conserved by virtue of a scale symmetry for a classical particle moving under the influence of an inverse-cube force law.

One can then use Newton's second law: $m\ddot{q} = -2k/q^3$, together with [32], to verify that $\frac{dN}{dt} = 0$.

**Classical particle simulations:** Here we use a stochastic model inversion approach to show that the generalised Lagrangian [16] can be used to detect scale invariance in timeseries. Specifically, we use Bayesian model inversion, followed by model reduction, to demonstrate face validity by using Dynamic Causal Modelling (DCM)[19] to distinguish between datasets that are known to be a) scale invariant; and b) scale free but not scale invariant — henceforth referred to simply as scale free.

We expand [16] to fifth order for a system with a time-independent Lagrangian and multiply the resultant expression by $q^\delta$, where $\delta$ is a constant, such that:

$$\mathcal{L} = q^{-\alpha+\delta} \sum_{y=0}^{4} C_y q^{y(\alpha-1)} \dot{q}^y, \qquad [33]$$

thus rendering $\delta$ a measure of deviation from scale invariance, i.e. we can use [33] to describe the two cases in which the system is a) scale invariant when $\delta = 0$; and b) scale free when $\delta \neq 0$. Note that we have now re-defined $\alpha$ as the exponent required for the Lagrangian to be scale invariant.

We then use the Euler-Lagrange equation [3] to recover the equation of motion associated with [33] which we use for all analyses presented, together with noise terms describing random, non-Markovian fluctuations[20] within the Statistical Parametric Mapping (SPM) software. In other words, we use [33] as a state space model of observable measurements $y$ by equipping the associated equations of motion with random fluctuations $\omega_f$ and mapping the (latent) states to observable quantities with additive observation noise $\omega_g$:

$$x = \dot{q} + \omega_f^{(x)}$$

$$\dot{x} = \frac{q^{-1} \sum_{y=0}^{4} C_y (1-y)((\alpha-1)y + \delta - \alpha) q^{(\alpha-1)y} \dot{q}^y}{\sum_{y=2}^{4} y(y-1) C_y q^{(\alpha-1)y} \dot{q}^{y-2}} + \omega_f^{(\dot{x})}$$

$$y = q + \omega_g. \tag{34}$$

This furnishes a dynamic causal model in the form of a stochastic differential equation (where the random fluctuations are assumed to be small). Crucially, the parameters $\theta_f = (\alpha, \delta, C_0, C_2, C_3, C_4)$ of this model can now be recovered from observations under the prior assumptions that the underlying dynamics take the form in [34]. The latter can be regarded as a *normal form for scale free systems* that become scale invariant when $\delta = 0$. This distinction affords the opportunity to assess the evidence for scale invariance by comparing models both with and without tight shrinkage priors on $\delta$. This is achieved using Bayesian model reduction; namely, comparing the evidence for models with and without constraints on $\delta$.

We use two ground truth datasets in the form of particle trajectories that are known a priori to be a) scale invariant, and b) scale free. In equations [28] through [30] we showed that the scale invariant case arises via a force that varies inversely as the cube of position, which in turn is known to result in an equiangular spiral trajectory[21] (Fig. 2A, left).

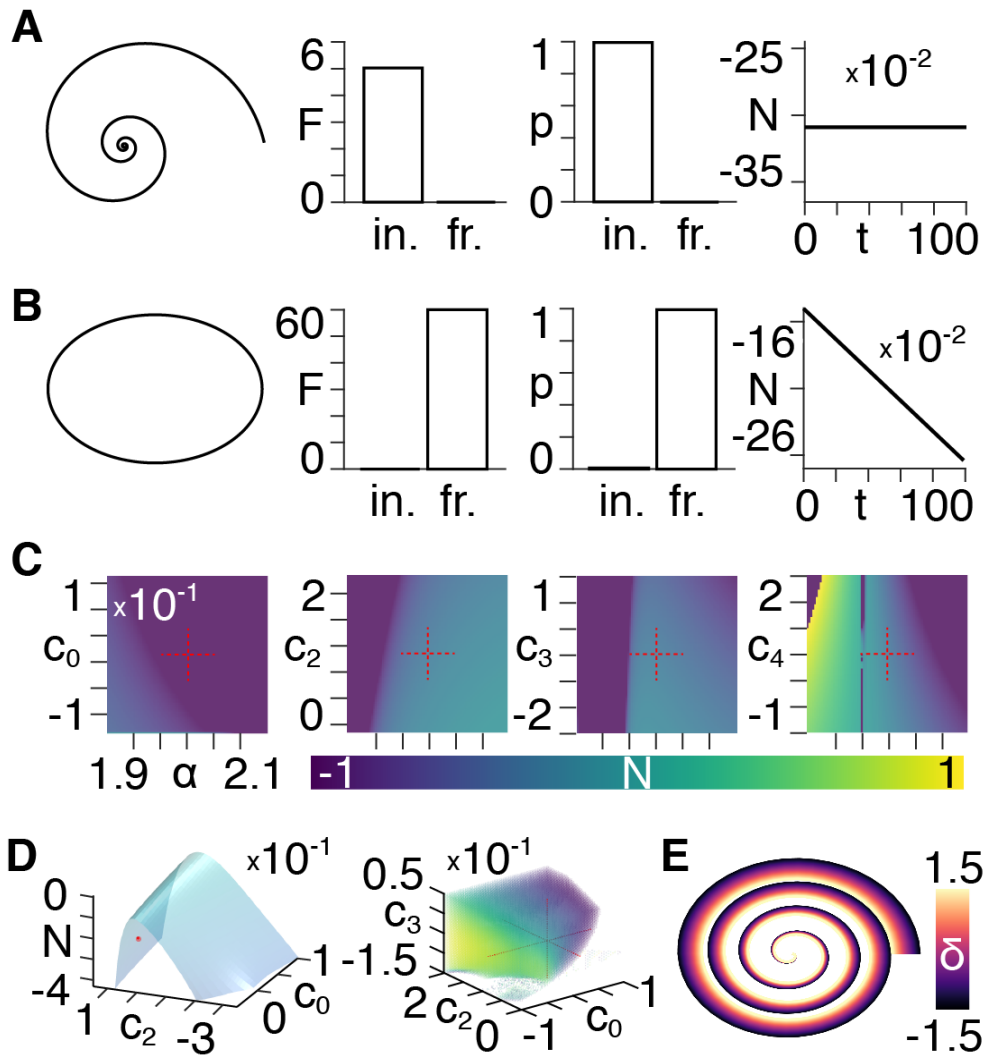

**Figure 2: simulations of a classical particle. A)** *In order from left to right: 1) the trajectory of a particle moving under the influence of a force that varies inversely as the cube of position; 2) Approximate lower bound log model evidence given by the free energy (F) following Bayesian model reduction for scale invariant (in.) and scale free (fr.) models; 3) Probabilities (p) derived from the log evidence; 4) Noether charge (N) as a function of time with low noise;* **B)** *Same layout as A) for a particle moving under the influence of a force that varies inversely as the square of position;* **C)** *All subsequent figures are shown with low levels of noise. Noether charge values between negative and positive unity, as indicated by the colour bar, for the four expansion coefficients (left to right) as a function of α. The centred red cross indicates the posterior densities in A);* **D)** *Left: Noether charge as a function of the first two expansion coefficients, with the posterior density values obtained in A) shown by the red dot; Right: Noether charge as a function of the first three expansion coefficients, with the posterior densities obtained from A) indicated by the centred red cross;* **E)** *The equation of motion resulting from a forward generative model for different values of δ as indicated by the colour bar.*

In the scale free case we use an inverse square force law which is known to result, for instance in planetary orbits, in an elliptical trajectory[22] (Fig. 2B, left). Note that we use a version of [34] in which we accommodate both an $x$ and $y$ coordinate, as shown in the accompanying code, to allow for the particles to trace 2-D trajectories.

We use Dynamic Expectation Maximisation (DEM)[23] to infer the latent states and estimate the above parameters (and hyperparameters; i.e. the precision components of random fluctuations on the states and observation noise). Having applied the optimization to the full model comprising a non-zero $\delta$ (i.e. scale free) in equation [33] we subsequently use Bayesian model reduction[24,25] to estimate the evidence for the reduced model in which $\delta = 0$ (i.e. scale invariant). We specify the reduced model by setting the prior variance over the $\delta$ parameter to zero, where $\delta$ is also given a prior mean of zero.

Using Bayesian model inversion, followed by model reduction, we show that the correct model is identified (Fig. 2A,B centre). We subsequently use the posterior expectations of the parameters for the full (scale free) and reduced (scale invariant) models show that the Noether charge is constant in time for the scale invariant (Fig. 2A, right), but not for the scale free case (Fig. 2B right).

We then explore the way in which the value of the Noether charge varies within the parameter space close to the posterior densities in terms of: $\alpha$ vs. each of the expansion coefficients (Fig. 2C); the first two expansion coefficients (Fig. 2D, left); and the first three expansion coefficients (Fig. 2D, right). Finally, we run the model forward to show the behaviour of the pure equation of motion in the $\delta$ parameter range $-1.5 < \delta < 1.5$, thus showing the transition from scale freeness with $\delta < 0$, through scale invariance ($\delta = 0$), and back to scale freeness with $\delta > 0$ (Fig. 2E).

**Neuroimaging data:** Here, we analyse murine calcium imaging[26] (rest and task) and macaque monkey fMRI[27] (rest and anaesthetised) datasets, using the same techniques as with the particle simulations described above. We show all results obtained for the resting states in Figure 3. Pre-processing of the murine calcium imaging[28] and macaque monkey fMRI[29] datasets were carried out as described previously.

The calcium imaging data were collected across an entire hemisphere of mouse cortex (Fig. 3A & B). We perform Bayesian model averaging across $n = 3$ mice with 10 trials of 10s (200 time points) duration each. We find that there is higher model evidence for scale invariance, as opposed to scale freeness, in a single region (Fig 3C, left). All other regions show either higher model evidence for scale freeness, or else cannot be statistically classified either as scale invariant or scale free (Fig 3C, right). No region emerges as scale invariant in the task state. We show a sample timecourse from the region classified as scale invariant, together with the estimated data following model inversion (Fig. 3D). We then run both the full (scale free) and reduced (scale invariant) models forward with low noise in the absence of external inputs in order to show the way in which the pure equations of motion evolve in time (Fig. 3E). We show the variational free energy (Fig. 3F) and associated probability (Fig. 3G) of the reduced model for the scale invariant region and run the model forward with parameters furnished by posterior densities from the scale invariant model to show that the Noether charge is constant in time (Fig. 3H).

In the macaque monkey fMRI data, we observed higher model evidence for scale invariance, as opposed to scale freeness, in a single cortical network (Fig. 3I). We show the variational free energy (Fig. 3J), associated probabilities (Fig. 2K), and Noether charge (Fig. 3L) for this scale invariant network. No network emerges as scale invariant in the anaesthetised state.

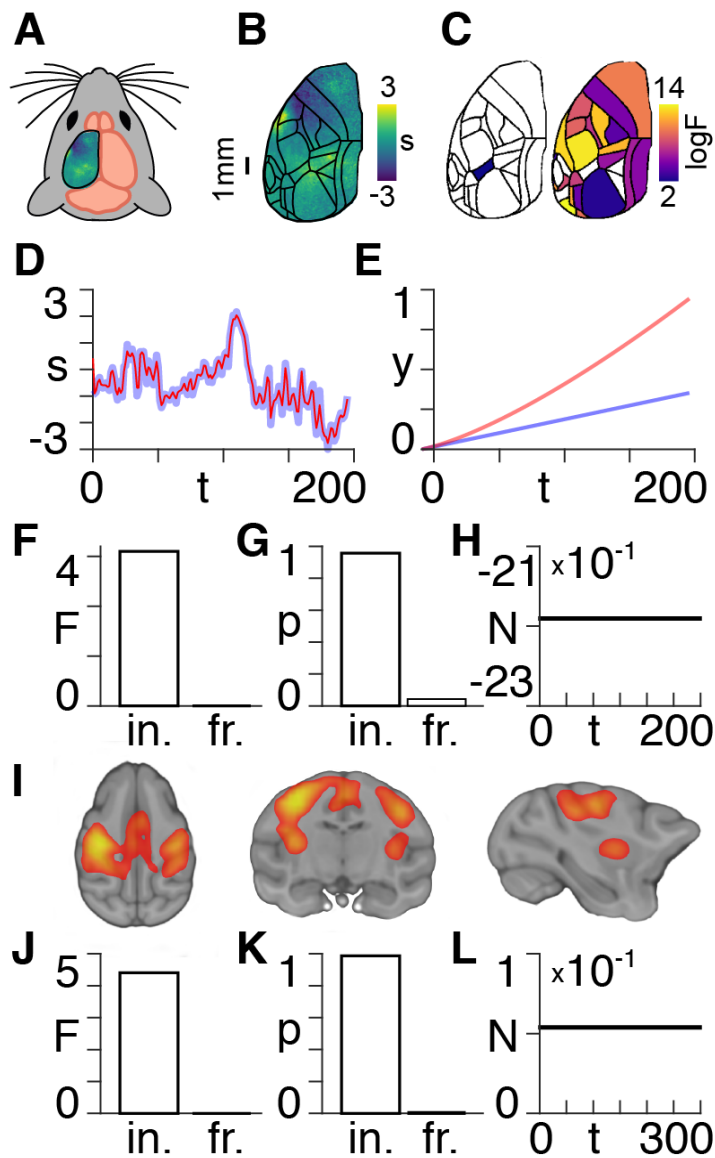

**Figure 3: Neuroimaging data A)** *Wide-field calcium imaging over the left hemisphere of a head-fixed mouse, expressing GCaMP6f in layer 2/3 excitatory neurons;* **B)** *Example z-scored (DF/F) activity averaged over a 10s trial length, shown as standard deviation (s) of the signal from the mean. Cortical areas are aligned to the Allen Mouse Common Coordinate Framework;* **C)** *Log variational free energy values corresponding to the colour bar, thresholded at $F = 3$ for regions found to have higher model evidence for scale invariance (left) and freeness (right);* **D)** *z-scored (DF/F) activity shown as standard deviation (s) of the signal from the mean from an example trial in one mouse in the scale invariant region (blue), together with the estimated timeseries following model inversion (red),* **E)** *Normalized timecourses of observable measurements (y) showing the evolution of the scale free (blue) and scale invariant (red) equations of motion with low noise and without driving inputs;* **F)** *Approximate lower bound log model evidence given by the free energy (F) following Bayesian model reduction for scale invariant (in.) and scale free (fr.) models in the calcium imaging data;* **G)** *Probabilities (p) derived from the log evidence in F;* **H)** *Noether charge (N) as a function of time for the calcium imaging data;* **I)** *The region explaining the highest amount of variance defined via temporal-concatenation probabilistic ICA, thresholded at $z > 3$;* **J)** *Approximate lower bound log model evidence given by the free energy (F) following Bayesian model reduction for scale invariant (in.) and scale free (fr.) models in the fMRI data;* **K)** *Probabilities (p) derived from the log evidence in J;* **L)** *Noether charge (N) as a function of time for the fMRI data.*

All methodological information pertaining to Figures 2 and 3, such as prior means, precisions, and posterior densities are made available here:

https://github.com/allavailablepubliccode/Symmetries

In contrast to the symmetries of translation in space, rotation in space, and translation in time, the known laws are not universally invariant under transformation of scale. In fact, as we showed in equations [28] to [30], the only way for a classical 1-D time-independent Lagrangian to qualify as scale invariant is if its potential energy term varies as the inverse square of position. More generally, we showed that a scale free dynamical system that follows the principle of stationary action is only scale invariant in the special case that its equation of motion scales inversely with time (see equation [12]). This restrictive condition may explain why symmetries under change of scale are not usually discussed in the context of dynamical systems.

Another reason could be that a symmetry is often defined as being contingent on a Lagrangian remaining invariant, which would only be possible in a scale transformation if the rescaling factors preceding the spatial and temporal variables cancelled each other in every term. However, such a definition of scale invariance would only be compatible with Noether's theorem if the Jacobian associated with the rescaling of the temporal variable were equal to unity. In the case of a non-unity Jacobian, quantities conserved by virtue of scale invariance only exist if one redefines what is meant by scale invariance to include a factor that cancels the Jacobian. No other definition leads to a conservation law. In other words, instead of satisfying the sufficient but not necessary condition of an invariant Lagrangian, we allow for the existence of scale symmetry via the sufficient and necessary condition of an invariant action. The point we wished to emphasize here is that, although only applicable under certain constraints, symmetries under change of scale can exist in real systems.

To demonstrate the practical applicability of the theoretical results, we derived an expression for a generalised scale invariant Lagrangian in the form of a power series expansion (see equation [16]) and showed that this can be used to distinguish scale invariance from scale freeness in ground-truth models of classical particle trajectories. We then used Noether's theorem to write the family of conservation laws that arise under change of scale for this generalised scale invariant Lagrangian.

As there is evidence indicating that neural systems operate with scale free dynamics[30-32] and also that they evolve via a stationary action principle[33-35] we used our methodology to analyse neural timeseries. Our objective was to demonstrate how scale invariance can be identified in real data and how the associated conserved quantities can be estimated. At this stage we did not seek to make any claims regarding the neurobiological properties of this conserved quantity. However, in future a more thorough exploration of scale invariance in neural systems using larger datasets across a wider range of neural states will be necessary.

When describing angular momentum one can turn to familiar real-world examples involving e.g. an ice skater spinning faster upon retracting her arms. Yet, if asked to provide a similarly intuitive understanding of the quantity conserved by virtue of scale invariance, we would be hard-pressed. We can, however, attempt to better understand this quantity by mapping the way in which it varies with respect to different parameters (see Fig. 2C-E). We can also make conjectures as to the nature of the said quantity in the context of neural systems, in which candidates for conservation laws have been empirically described; e.g. in the balance of excitation/inhibition[36,37].

In summary, we derived a generalised equation of motion that can be used to estimate scale invariance – as well as the associated conserved quantities – in timeseries from any scale free dynamical system following the principle of stationary action.

**Acknowledgements**

E.D.F. and R.L. were funded by the Medical Research Council (Ref: MR/R005370/1). K.J.F. was funded by a Wellcome Principal Research Fellowship (Ref: 088130/Z/09/Z). R.J.M. was funded by the Wellcome/EPSRC Centre for Medical Engineering (Ref: WT 203148/Z/16/Z).

**Author contributions**

Y.G.S. and F.H. collected the murine calcium imaging data; All authors designed and performed research, analysed data and wrote the paper.

**Competing interests**

The authors declare no competing interests.